\documentclass[%
twocolumn,
amsmath,amssymb,
aps,
]{revtex4}

\usepackage{graphicx}
\usepackage[colorlinks=true,linkcolor=blue,anchorcolor=blue, citecolor=blue,urlcolor=blue]{hyperref}

\begin{document}
\preprint{APS/123-QED}
\title{Rotational symmetry breaking and partial Majorana corner states in a high-T$_c$ superconductor based heterostructure}
\author{Yu-Xuan Li}
\author{Tao Zhou}%
\email{tzhou@scnu.edu.cn}
\affiliation{Guangdong Provincial Key Laboratory of Quantum Engineering and Quantum Materials, GPETR Center for Quantum Precision Measurement, SPTE,
	and
	Frontier Research Institute for Physics,
	South China Normal University, Guangzhou 510006, China
}
\begin{abstract}
Applying a microscopic model,
 we study theoretically the quasiparticle excitation of a two-dimensional topological insulator (TI) being in proximity to a high-T$_c$ superconductor.
In the momentum space, the proximity induced pairing term in the TI layer includes both the singlet channel and triplet channel, leading to the
$\mathcal{C}_4$ rotational symmetry breaking of the energy bands and the quasiparticle spectra.
For a cylinder geometry, the zero energy edge states may appear but they are localized at the upper boundary. For the finite-size system with open boundaries,
the zero energy states at the upper boundary disappear and the Majorana bound states emerge at the two lower corners.
All of the results can be understood well through exploring the pairing order parameter and the anomalous Green's function.

\end{abstract}
\maketitle



Higher order topological materials have been studied intensively in the past several years~\cite{re32,re9,re19,re21,re190,RN4,re34,re100,re35,re118,re37,re130,re117,re122,re121,re2}.
For a $d$-dimensional higher order topological material, the system is fully gapped at its ($d-1$)-dimensional boundaries. Topological protected gapless states emerge at even lower dimensional boundaries.
Recently much attention has been paid on the two-dimensional (2D) higher order topological superconductor, namely, the system is gapped at its one-dimensional edge and hosts Majorana bound states at the corners.
High-T$_c$ superconductors have higher superconducting transition temperature and larger pairing gap in comparison with conventional superconductors. It is natural to make use of high-T$_c$ superconductors
to obtain effective topological superconductors~\cite{re126,re82,re83,re1,re80,re81,re8}.
Very recently, many efforts have been made to realize the 2D higher order topological superconductor or Majorana corner states with the high-T$_c$ superconductor based heterostructures~\cite{re35,RN4,re34,re100,re118,re37}.

In a heterostructure system including two layers or more, the superconducting pairing term will be induced from one layer to another, known as the proximity effect.
Previously, when studying the higher order topology in a certain heterostructure system, the proximity effect was usually phenomenologically considered with adding the effective pairing term directly and neglecting the mixing of the band structures from different layers~\cite{re35,RN4,re34,re100,re118,re37,re130,re117,re122}.
While actually, for a hybrid system, a more microscopic model should include the original Hamiltonian describing different materials and consider their coupling. The effective pairing term in a non-superconducting material is induced by the tunneling of the systems.
Such microscopic model has indeed been considered in some first order topological system~\cite{re106,re108,re8,re124,re125,proximity,re128}.
Some interesting properties emerge due to the mixing of the band structures. Especially,
the induced pairing symmetry is not necessarily identical to the original one of the superconductor~\cite{re8,proximity}. Therefore,
it may be fundamental and of importance to consider a
microscopic model when exploring the higher order topology in certain heterostructure system.

In this paper, we study the quasiparticle excitation in a previously proposed higher order topological superconductor candidate system~\cite{re100,RN4,re34}, with considering
 a 2D topological insulator (TI) coupled with a $d$-wave high-T$_c$ superconductor.
With a microscopic model, our results indicate that this system is different from the conventional first order topological superconductor and the higher order one.
In the momentum space or the system bulk, the quasiparticle spectra in the 2D TI layer are fully gapped. The $\mathcal{C}_4$ rotational symmetry of the band structure and the spectral function is broken by the proximity induced pairing term.
With a cylinder geometry, along one boundary the system is gapless while along other boundaries the system is fully gapped.
In the real space with finite system size, the Majorana bound states emerge at partial corners. These results can be explained soundly through exploring the pairing order parameter and
the anomalous Green's function of the 2D TI layer.

We start from the Hamiltonian considering the coupling of a 2D TI and a $d$-wave superconductor, with $H =H_{\mathrm{TI}}+H_{\mathrm{SC}}+H_{\mathrm{I}}$.
$H_{TI}$ describes a 2D TI in the square lattice~\cite{re200,re201},
\begin{equation}
\begin{aligned}
H_{TI} &=\sum_{{\bf k}}C_{{\bf k}}^\dagger (h_\mathbf{k}\sigma_3s_0+2\lambda_{0}\sin k_x\sigma_1s_3\\&+2\lambda_0\sin k_y\sigma_2 s_0)C_{{\bf k}},
\end{aligned}
\end{equation}
with $h_\mathbf{k}=h_0-2t(\cos k_x+\cos k_y)$. $C^\dagger_{\bf k}$ is a forth order wave vector with $C^\dagger_{\bf k}=(c^\dagger_{{\bf k}1\uparrow},c^\dagger_{{\bf k}2\uparrow},c^\dagger_{{\bf k}1\downarrow},
c^\dagger_{{\bf k}2\downarrow})$. The subscripts $1,2$ and $\uparrow$, $\downarrow$ represent
the orbital and spin indices, respectively. $s_i$ and $\sigma_i$ are identity matrix $(i=0)$ or Pauli matrices $(i=1,2,3)$ in the spin and orbital spaces, respectively.

$H_{SC}$ describes a $d$-wave superconductor,
\begin{equation}
H_{SC}=\sum_{{\bf k}\sigma}\varepsilon_{\bf k}d^\dagger_{{\bf k}\sigma}d_{{\bf k}\sigma}+\sum_{{\bf k}}\Delta_{\bf k}(d^\dagger_{{\bf k}\uparrow}d^\dagger_{{-\bf k}\downarrow}+h.c.),
\end{equation}
with $\varepsilon_{\bf k}=-2t(\cos k_x+\cos k_y)-\mu$, and $\Delta_{\bf k}=2\Delta_0 (\cos k_x-\cos k_y)$.

$H_I$ is the interlayer single particle hopping term, with
\begin{equation}
H_{I}=-t_\perp \sum_{{\bf k}\alpha\sigma}(c_{{\bf k}\alpha\sigma}^\dagger d_{{\bf k}\sigma}+h.c.).
\end{equation}

To study the edge states and the Majorana corner states numerically,
let us transform
the Hamiltonian to the real space
by preforming a Fourier transformation. Then the Hamiltonian is rewritten as,
\begin{equation}
\begin{aligned}
H_{TI} &=-t\sum_{{\bf i}\alpha}(C_{{\bf i}}^\dagger \sigma_3s_0   C_{{\bf i}+\alpha}+h.c.)+h_0\sum_{{\bf i}\alpha}C_{{\bf i}}^\dagger \sigma_3s_0   C_{{\bf i}}
\\&-\lambda_0\sum_{{\bf i}}(iC_{{\bf i}}^\dagger\sigma_1s_3C_{{\bf i}+\hat{x}}-iC_{{\bf i}}^\dagger\sigma_2s_0C_{{\bf i}+\hat{y}}+h.c.),
\end{aligned}
\end{equation}
\begin{equation}
\begin{aligned}
H_{sc} &=-t\sum_{{\bf i}\alpha\sigma}(d_{{\bf i}\sigma}^\dagger d_{{\bf i}+\alpha,\sigma}+h.c.)-\mu\sum_{{\bf i}\sigma}d_{{\bf i}\sigma}^\dagger d_{{\bf i}\sigma}
\\&+\sum_{\langle{\bf ij}\rangle}(\Delta_{\bf ij}d_{{\bf i}\uparrow}^\dagger d_{{\bf j}\downarrow}^\dagger+h.c.),
\end{aligned}
\end{equation}
and
\begin{equation}
H_{I}=-t_\perp \sum_{{\bf i}\alpha\sigma}(c_{{\bf i}\alpha\sigma}^\dagger d_{{\bf i}\sigma}+h.c.),
\end{equation}
For the $d$-wave pairing, the site ${\bf j}$ is the nearest-neighbor site to the site ${\bf i}$, with
$\Delta_{\bf ij}=\pm\Delta_0$ ($\pm$ depends on $\langle{\bf ij}\rangle$ along the $x$ direction or $y$ direction).

The total Hamiltonian can be rewritten as the matrix form.
In the momentum space, the matrix is a $12\times 12$ matrix $\hat{M}_{\bf k}$ with
$H=\sum_{\bf k}\Psi^{\dagger}_{\bf k}\hat{M_{\bf k}}\Psi_{\bf k}$. The vector $\Psi^{\dagger}_{\bf k}$ is expressed as,
\begin{equation}
\Psi^\dagger_{\bf k}=(C^\dagger_{\bf k},C_{-{\bf k}},d^\dagger_{{\bf k}\uparrow},d^\dagger_{{\bf k}\downarrow},d_{-{\bf k}\uparrow},d_{-{\bf k}\downarrow}).
\end{equation}
In the real space, the Hamiltonian matrix $\hat{M}$ is a $N\times N$ lattice ($H=\Psi^{\dagger}\hat{M}\Psi$),
with $N=8N_1+4N_2$ ($N_1$ and $N_2$ are the number of sites in the 2D TI layer and the $d$-wave
superconducting layer, respectively). The vector $\Psi^{\dagger}$ is expressed as,
\begin{eqnarray}
\Psi^\dagger =(C^\dagger_1,C_{1},\cdots,C^\dagger_{N_1},C_{N_1},d^\dagger_{1\uparrow},\cdots,d_{N_2\downarrow}).
\end{eqnarray}
Here the vector $C^\dagger_{i}$ is the Fourier transformation of the vector $C^\dagger_{\bf k}$

Diagonalizing the Hamiltonian matrix, we obtain the retarded Green's Function matrix $\hat G$, with the elements being expressed as,
\begin{equation}
G_{ij}(E)=\sum_n\frac{u_{in}u^{*}_{jn}}{E-E_n+i\Gamma}.
\end{equation}
Here $u_{in}$ and $E_n$ are the eigen-vectors and eigen-values of the Hamiltonian matrix, respectively.

In the momentum space,
the spectral function of the 2D TI layer is calculated through,
\begin{equation}
A({\bf k},E)=-\frac{1}{\pi}\sum^4_{p=1}\mathrm{Im} G_{pp}({\bf k},E).
\end{equation}

The proximity induced pairing term for the orbital $\tau$ can be studied through the mean-field pairing order parameter, expressed as,
\begin{equation}
\Delta_\tau({\bf k})=\langle c^\dagger_{{\bf k}\tau\uparrow}c^\dagger_{-{\bf k}\tau\downarrow}\rangle=\sum_n{u^{*}_{\tau,n}({\bf k})u_{\tau+6,n}({\bf k})}f(E_n),
\end{equation}
where $f(x)$ is Fermi distribution function. 

In the real space,
the effective pairing order parameter of the sites $i$ and $j$ for the orbital $\tau$ can be expressed as,
\begin{equation}
\Delta^\tau_{ij}=\sum_nu^{*}_{h(i),n}u_{h(j)+6,n}f(E_n),
\end{equation}
with $h(i)=\tau+8(i-1)$.

We define the site-dependent $d$-wave pairing magnitude for the orbital $\tau$, with
\begin{equation}
\Delta^\tau_i=\mid \Delta^\tau_{i,i+\hat{x}}+\Delta^\tau_{i,i-\hat{x}}
-\Delta^\tau_{i,i+\hat{y}}-\Delta^\tau_{i,i-\hat{y}}\mid.
\end{equation}
At the system edge, $\Delta^\tau_i$ is expressed as
\begin{equation}
\Delta^\tau_i=\mid 2(\Delta^\tau_{i,i+\hat{\alpha}}+\Delta^\tau_{i,i-\hat{\alpha}}) \mid  \qquad (\alpha=x,y).
\end{equation}

  \begin{figure}
\centering
  \includegraphics[width=1.8in]{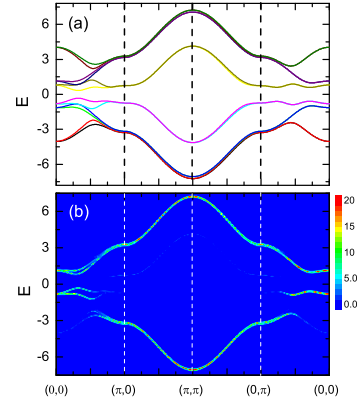}
\caption{(Color online) (a)The energy bands in the momentum space.
(b) The spectral function as functions of the energy and the momentum in the momentum space.
}\label{fig1}
\end{figure}

The local density of states (LDOS) of the site $i$ in the 2D TI layer can be calculated through the Green's function in the real space, with,
\begin{equation}
\rho_i(E)=-\frac{1}{\pi}\sum^4_{p=1}\mathrm{Im} G_{m+p,m+p}(E).
\end{equation}
with $m=8(i-1)$.

In the following presented results, the parameters are set as $t=1$, $\lambda_0=0.5$, $h_0=3$, $\mu=-0.3$, $\Delta_0=0.2$, $t_\perp=0.8$, and $\Gamma=0.01$. Our main results are not sensitive to the parameters we considered.


We first study the energy spectrum in the system bulk. The energy bands obtained by diagonalizing the $12\times12$ Hamiltonian in the momentum space are presented in Fig.~1(a). One striking feature is
that the $\mathcal{C}_4$ rotational symmetry of the energy bands is broken, namely, the quasiparticle bands are significantly different along the $(k_x,0)$ direction and the $(0,k_y)$ direction.
However, such a rotational symmetry is preserved by adding the $d$-wave pairing term directly into the 2D TI Hamiltonian and neglecting the inter-layer hopping~\cite{re100,RN4,re34}.
Experimentally, the energy bands can be detected through the spectral function. Here for the 2D TI layer, we have checked numerically that the spectra are fully gapped in the whole Brillouin zone.
The spectral function of the 2D TI layer as functions of the momentum and the energy is presented in Fig.~1(b).
Also, here the $\mathcal{C}_4$ symmetry is broken and such asymmetrical behavior may be detected by later experiments.

  \begin{figure}
\centering
  \includegraphics[width=3in]{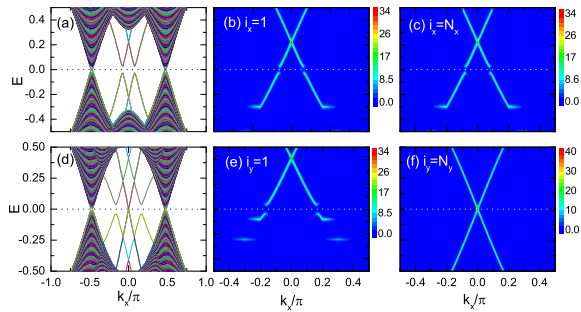}
\caption{(Color online) Numerical results considering a cylinder geometry. (a) The eigenvalues
of the Hamiltonian with the open boundary condition along the $x$-direction being considered.
(b) The spectral function at the $i_x=1$ boundary. (c) The spectral function at the $i_x=N_x$ boundary.
(d-f) Similar to panels (a-c) but for the open boundary condition along the $y$-direction.
}\label{fig2}
\end{figure}

We now study the edge states through considering the cylinder geometry with the open boundary condition along one direction and the periodic boundary condition along the other.
The energy bands with the open boundary condition along the $x$ direction
are presented in Fig.~2(a). The whole energy bands should include both the bulk states and the edge states. Note that, at these two boundaries, the energy bands are still fully gapped.
The gapped edge states can be studied more clearly through the spectral functions at the two system boundaries, which are presented in Figs.~2(b) and 2(c), respectively. As is displayed, the spectra at these two boundaries are exactly the same. An energy gap
around 0.05 is revealed. This gapped behavior originated from the proximity induced effective $d$-wave pairing term~\cite{re100,RN4,re34}.

We turn to discuss the edge states when the open boundary rotates to the $y$ direction.
The corresponding energy bands and the spectral functions at the system boundaries ($i_y=1$ and $i_y=N_y$) are presented in Figs.~2(d)-2(f).
As presented, both the energy bands and the spectral functions are different from those shown in Figs.~2(a)-2(c), indicating the $\mathcal{C}_4$ symmetry breaking. Interestingly, here from the energy band spectra [Fig.~2(a)], the edge states are gapless, significantly different from previous theoretical results~\cite{re100,RN4,re34}.
The spectral functions at the two boundaries are also different, namely, at the $i_y=1$ boundary, the spectrum is still gapped, with the gap magnitude around 0.08.
This value is even larger than those at the $i_x=1$ and $i_x=N_x$ boundaries. At the $i_y=N_y$ boundary, the edge state is gapless.
Thus for the cylinder system, the four boundaries are nonequivalent and the zero energy states appear at the $i_y=N_y$ boundary. Such asymmetric behavior may be detected by later experiments.

  \begin{figure}
\centering
  \includegraphics[width=3in]{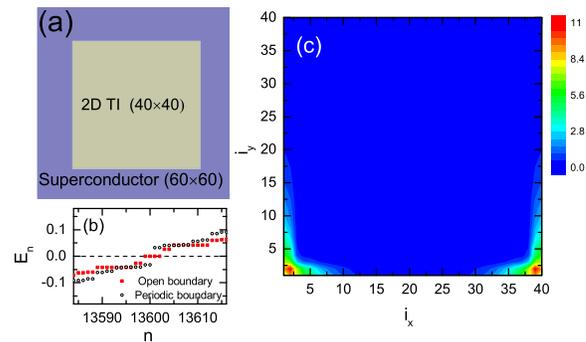}
\caption{(Color online) (a) Schematic illustration of a 2D TI being grown on a d-wave high-T$_c$ superconductor. (b) The eigenvalues of the Hamiltonian in the real space. (c) The intensity plot of the zero energy LDOS in the real space.
}\label{fig3}
\end{figure}

Now let us discuss the possible Majorana corner states.
We consider a 2D TI (with a finite-size lattice $40\times40$) being placed on a larger high-T$_c$ superconductor, as sketched in Fig.~3(a).
Considering open boundaries along both $x$ and $y$ directions, the Hamiltonian matrix is diagonalized. The corresponding eigenvalues are presented in Fig.~3(b).
Four zero energy eigenvalues are revealed. Note that there is no zero energy eigenvalue when periodic boundary conditions are taken.
Therefore, these four eigenvalues should locate at the system edges, relating to Majorana zero modes. For the topological superconducting system,
four zero energy eigenvalues should come from two zero energy physical quasiparticles, corresponding to two pairs of Majorana zero modes at the system boundaries.
The distributions of the two pairs of Majorana zero modes can be seen from the zero energy LDOS, as displayed in Fig.~3(c).
These two pairs locate at two lower corners of the system (one pair at each corner). As for the upper corners, no zero energy state exists.

The asymmetric results presented above can
be understood through exploring the pairing order parameter in the 2D TI layer. The order parameter for the orbital 1
 as a function of the momentum ${\bf k}$ is presented in Fig.~4(a). As is seen, here both the $\mathcal{C}_4$ symmetry and the inversion symmetry for the magnitude of the order parameter are broken. We can separate the whole pairing order parameter as the singlet channel $\Delta_e$ and the triplet channel $\Delta_o$ [$\Delta_1({\bf k})=\Delta_e({\bf k})+\Delta_o({\bf k})]$, with $\Delta_e({\bf k})=1/2[\Delta_1({\bf k})+\Delta_1(-{\bf k})]$ and $\Delta_o({\bf k})=1/2[\Delta_1({\bf k})-\Delta_1(-{\bf k})]$. The corresponding numerical results for these two channels are displayed in Figs.~4(b) and 4(c), respectively. For the singlet channel, the result is consistent with the $d_{x^2-y^2}$ pairing symmetry. For the triplet channel, a $p$-wave symmetry is revealed. The whole effective pairing symmetry should be $d$+$p$ wave. As a result, both the $\mathcal{C}_4$ symmetry and the inversion symmetry are broken for the spin-dependent spectral function for the orbital 1. As to the whole spectral function, the $\mathcal{C}_4$ symmetry is broken, while the inversion symmetry preserves due to the time reversal symmetry of the system.

  \begin{figure}
\centering
  \includegraphics[width=3.3in]{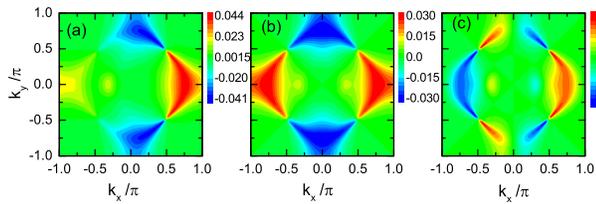}
\caption{(Color online) Intensity plots of the pairing order parameter of the 2D TI layer.
(a) The whole pairing order parameter. (b) The pairing order parameter in the singlet channel part. (c) The pairing order parameter in the triplet channel part.
}\label{fig4}
\end{figure}

In the real space with open boundaries along both the $x$ direction and the $y$ direction, the site dependent $d$-wave pairing order parameters $\Delta_i$ for the orbital 1 are displayed in Fig.~5(a). The two dimensional cuts of $\Delta_i$ along the four boundaries are plotted in Fig.~5(b).
As is seen, at the $i_y=1$ boundary, the magnitude is relatively large.
For the $i_x=1$ and $i_x=N_x$ boundaries , the same magnitudes are revealed while they are smaller than that at the $i_y=1$ boundary.
As for the $i_y=N_y$ boundary, the magnitude of the induced pairing order parameter is rather small.
Therefore, with a cylinder geometry, the topological protected gapless edge states may survive at this boundary.
 These results are well consistent with the numerical results of the spectral function at the system boundaries [Fig.~2]. In the mean time, for the finite-size system with open boundaries, the Majorana corner states will generally emerge at the corners where the sign of the $d$-wave pairing term changes sign~\cite{re100,RN4,re34}. However,
since the effective $d$-wave order at the $i_y=N_y$ boundary is too small, the upper Majorana corner states will not come into being. Therefore, here the Majorana corner states only emerge at lower boundaries, as presented in Fig.~3.

\begin{figure}
\centering
\includegraphics[width=3in]{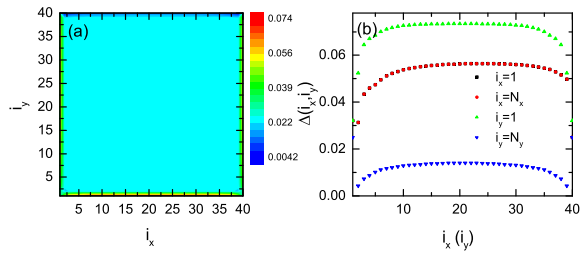}
\caption{(Color online) (a) The intensity plot of the site-dependent $d$-wave pairing order parameter of the 2D TI layer.
(b) The replot of the $d$-wave pairing order parameter at the four boundaries.
}\label{fig5}
\end{figure}

The effective quasiparticle pairing in the 2D TI layer can also be explored through the anomalous Green's function~\cite{supp}.
In the momentum space, our analytical results of the anomalous Green's function verify that the singlet channel and the triplet channel pairing indeed coexist and leading to the $\mathcal{C}_4$ symmetry breaking. Numerically, the imaginary parts of the anomalous Green's function are consistent with the pairing order parameter shown in Fig.~4.

The asymmetric behavior at the system boundaries can also be well understood through the anomalous Green's function and the effective Hamiltonian at the system boundaries~\cite{supp}.
 Without coupling to the superconductor, the 2D TI is gapless at the boundaries, with the linear quasiparticle dispersion crossing the Fermi energy.
 The effective Hamiltonian at boundaries should include the intraorbital hopping term and the interorbital hopping term.
 When coupling to a superconductor, both terms contribute to the effective pairing at the boundaries.
 For the $i_y=N_y$ boundary, the interorbital hopping constant is opposite to the intraorbital one. These two contributions cancel out. Therefore, the effective pairing magnitude at this boundary is rather small.
For the $i_y=1$ boundary, these two contributions add up directly. The pairing magnitude at this boundary is relatively large. For the $i_x=1$ or $i_x=N_x$ boundary,
the interorbital hopping constant is imaginary.
An effective $d$-wave pairing term is also induced, while its magnitude is relatively smaller
than that of the $i_y=1$ boundary. The analytical results of the anomalous Green's function at these four boundaries are consistent with the above discussions.

At last, we would like to make several remarks. First, previously the proximity effect has been widely used to artificially create topological superconductors. Very recently, based on the proximity effect, several proposals have also been proposed to realize the second order topological superconductor in various hybrid systems. At this stage, studying the proximity effect in a more strict way, especially for the possible higher order topological superconducting system, is timely and of broad interest. Secondly, for the finite-size system, the Majorana bound state only emerge at lower corners. The zero energy degeneracy is reduced to half of those obtained from previous phenomenological theoretical results. The reduced degeneracy may make the Majorana bound states be more controllable. Thirdly, Our main results can be well understood. In the momentum space, the proximity induced pairings include both the $d$-wave component and the $p$-wave component, leading to the breaking of $\mathcal{C}_4$ symmetry.
At system boundaries in the real space, the proximity induced pairing terms are mainly contributed by the edge states of the TI.
The edge states are contributed by the intraorbital channel and the interorbital channel. The coexistence of these two channels may strengthen or weaken the induced pairing gap, leading to the interesting asymmetric behaviours.
At last, it is needed to pinpoint that the present system is different from both the first order topological superconductor and the higher order one, namely,
the system can be gapless for a certain one-dimensional boundary while it is fully gapped for other boundaries. In the meantime, the Majorana bound states emerge at partial corners of the finite system. It is also interesting that the zero energy states may shift from the upper boundary to the lower corners when the boundary condition along the $x$-direction changes.

\appendix
\section{Derivation of the anomalous Green's function}
\subsection{Anomalous Green's function in the momentum space}
The effective pairing term in the orbital $\tau$ of the 2D TI layer can also be described by the anomalous retarded Green's function, expressed as $F_\tau({\bf k},\omega)=\langle\langle c_{{\bf k}\tau\uparrow}^\dagger|c_{-{\bf k}\tau\downarrow}^\dagger\rangle\rangle_{\omega+i\Gamma}$.
 It can be calculated numerically from the off-diagonal components of the retarded Green's functions [Eq.(9) in the main text], with
\begin{equation}
F_\tau({\bf k},\omega)=\sum_n\frac{{u^{*}_{\tau,n}({\bf k})u_{\tau+6,n}({\bf k})}}{\omega-E_n+i\Gamma}.
\end{equation}
We define a momentum dependent function $F_\tau({\bf k})$ to describe the effective pairing symmetry, with
\begin{equation}
F_\tau({\bf k})=-\int_{-\infty}^0\mathrm{Im}F_\tau({\bf k},\omega)d\omega.
\end{equation}
$F_\tau({\bf k})$ is generally proportional to the pairing order parameter $\Delta_\tau({\bf k})$.

The Green's function can also be calculated analytically based on the equation of motion,
\begin{equation}
\omega\langle\langle A|B\rangle\rangle_\omega=\langle\left[A,B\right]_{+}\rangle+\langle\langle\left[A,H\right]|B\rangle\rangle_\omega.\label{eqmo}
\end{equation}
Repeated application of Eq. (\ref{eqmo}) to the higher-order
Green's functions on the right hand side $\langle\langle\left[A,H\right]|B\rangle\rangle_\omega$,
will result in a set of solvable equations.

Let us define the following Green's function elements, with $x_1=\langle\langle c_{{\bf k}1\uparrow}^\dagger|c_{-{\bf k}1\downarrow}^\dagger\rangle\rangle_\omega$,
$x_2 = \langle\langle c_{{\bf k}2\uparrow}^\dagger|c_{-{\bf k}1\downarrow}^\dagger\rangle\rangle_\omega$,
$x_3 = \langle\langle d_{{\bf k}\uparrow}^\dagger|c_{-{\bf k}1\downarrow}^\dagger\rangle\rangle_\omega$,
$x_4 = \langle\langle d_{-{\bf k}\downarrow}|c_{-{\bf k}1\downarrow}^\dagger\rangle\rangle_\omega$,
$x_5 = \langle\langle c_{-{\bf k}1\downarrow}|c_{-{\bf k}1\downarrow}^\dagger\rangle\rangle_\omega$,
$x_6 = \langle\langle c_{-{\bf k}2\downarrow}|c_{-{\bf k}1\downarrow}^\dagger\rangle\rangle_\omega$. Repeated application of Eq. (\ref{eqmo}) on these six elements, we have the following equations,
\begin{subequations}
	\begin{eqnarray}
	ax_1&=&fx_2+t_\perp x_3
	\\
	bx_2&=&ex_1+t_\perp x_3
	\\
	cx_3&=&gx_4+t_\perp (x_1+x_2)
	\\
	dx_4&=&gx_3-t_\perp (x_5+x_6)
	\\
	bx_5&=&1-fx_6-t_\perp x_4
	\\
	ax_6&=&-ex_5-t_\perp x_4,
	\end{eqnarray}
\end{subequations}
with $a=\omega-M(\mathbf{k})$, $b=\omega + M(\mathbf{k})$, $c=\omega-m(\mathbf{k})$,
$d=\omega+m(\mathbf{k})$, $e=2\lambda_0\sin(k_x)-2i \lambda_0\sin(k_y)$, $f=2\lambda_0\sin(k_x) + 2i\lambda_0\sin(k_y)$,
and $g = \Delta_{\bf k}$.

The anomalous Green's function of orbit 1 is then obtained by solving the equations,
\begin{equation}
\langle\langle c_{{\bf k}1\uparrow}^\dagger|c_{-{\bf k}1\downarrow}^\dagger\rangle\rangle=\frac{g t_\perp ^2 (a-e) (b+f)}{\Omega}=\frac{gt_\perp^2(C_{even}+C_{odd})}{\Omega},\label{anlyagf}
\end{equation}
with $\Omega=-\left(c d-g^2\right) (a b-e f)^2+t_\perp ^2 (a b-e f) (a (c+d)+b (c+d)-(c-d) (e+f))+t_\perp ^4 (-(a+b-e-f)) (a+b+e+f)$, $C_{even}=\omega^2-M^2(\mathbf{k})-4\lambda_0(\sin^2(k_x)+\sin^2(k_y))$, $C_{odd}=4\omega\lambda i\sin(k_y)-4M(\mathbf{k})\lambda_0\sin(k_x)$. As is seen, the anomalous Green's function indeed includes an even channel and an odd channel. The $\mathcal{C}_4$ rotational is broken.

With the analytical extension ($\omega\rightarrow\omega+i\Gamma$), the retarded Green's function is obtained. Then the momentum dependent function $F_1({\bf k})$ can be calculated.
We have checked numerically that the results are qualitatively consistent with the order parameter shown in Fig.~\ref{fig4}.

\subsection{Anomalous Green's function at the system boundaries}
Now let us discuss the anomalous Green's function at the four system boundaries. Generally gapless modes appear at the boundaries of a 2D TI.
We can rewrite the effective Hamiltonian at a certain boundary of 2D TI. For the case of the $i_y=1$ boundary, the effective Hamiltonian is written approximately as,
\begin{equation}
H_{(i_y=1)}=-\sum_{\sigma,\alpha}\sigma k c_{k\alpha\sigma}^\dagger c_{k\alpha\sigma}-\sum_\sigma(\sigma k c_{k1\sigma}^\dagger c_{k2\sigma}+h.c),
\end{equation}
with $\sigma$ taking $\pm1$ for the spin up and spin down quasiparticles, respectively.

The Hamiltonian for the superconducting layer is written as,
\begin{equation}
	H_{SC}=\sum_\sigma\epsilon_{k} d_{k\sigma}^\dagger d_{k\sigma}+(\Delta_{ k} d_{k\uparrow}^\dagger d_{-k\downarrow}^\dagger+h.c).
\end{equation}
And the interlayer coupling is expressed as,
\begin{equation}
	H_I=-\sum_{\sigma,\alpha}(t_\perp c_{k\alpha\sigma}^\dagger d_{k\sigma}+h.c).
\end{equation}

Defining the following six Green's function elements, with
$x_1=\langle\langle c_{k1\uparrow}^\dagger|c_{-k1\downarrow}^\dagger$$\rangle$$\rangle_\omega$, $x_2=\langle\langle c_{k2\uparrow}^\dagger|c_{-k1\downarrow}^\dagger$$\rangle$$\rangle_\omega$, $x_3=\langle\langle d_{k\uparrow}^\dagger|c_{-k1\downarrow}^\dagger\rangle\rangle_\omega$, $x_4=\langle\langle d_{-k\downarrow}|c_{-k1\downarrow}^\dagger\rangle\rangle_\omega$, $x_5=\langle\langle c_{-k1\downarrow}|c_{-k1\downarrow}^\dagger\rangle\rangle_\omega$, $x_6=\langle\langle c_{-k2\downarrow}|c_{-k1\downarrow}^\dagger\rangle\rangle_\omega$. According to the equation of motion, we have
\begin{subequations}
	\begin{eqnarray}
	(\omega+k)x_1&=&-kx_2-tx_3\\
	(\omega+k)x_2&=&-kx_1-tx_3\\
	(\omega-\epsilon_{k})x_3&=&-t(x_1+x_2)+\Delta_{k} x_4\\
	(\omega+\epsilon_{k})x_4&=&t(x_5+x_6)+\Delta_{k} x_3\\
	(\omega+k)x_5&=&1-k x_6+tx4\\
	(\omega+k)x_6&=&-k x_5+tx_4
	\end{eqnarray}
\end{subequations}
Solving the above equations, the anomalous Green's function for the orbital 1 at $y=1$ boundary is obtained,
\begin{equation}
F_{(i_y=1)}=\frac{\Delta_{k} t_\perp^2}{\Delta_k^2(2k+\omega)^2+\epsilon_k^2(2k+\omega)^2-(-2t_\perp^2+\omega(2k+\omega))^2}.
\end{equation}

The effective Hamiltonian at other boundaries can be written as,
\begin{equation}
H_{(i_y=N_y)}=\sum_{\sigma,\alpha}\sigma k c_{k\alpha\sigma}^\dagger c_{k\alpha\sigma}-\sum_\sigma(\sigma k c_{k1\sigma}^\dagger c_{k2\sigma}+h.c),
\end{equation}

\begin{equation}
H_{(i_x=1)}=\sum_{\sigma,\alpha}\sigma k c_{k\alpha\sigma}^\dagger c_{k\alpha\sigma}-\sum_\sigma(i k c_{k1\sigma}^\dagger c_{k2\sigma}+h.c),
\end{equation}
and
\begin{equation}
H_{(i_x=N_x)}=-\sum_{\sigma,\alpha}\sigma k c_{k\alpha\sigma}^\dagger c_{k\alpha\sigma}-\sum_\sigma(i k c_{k1\sigma}^\dagger c_{k2\sigma}+h.c),
\end{equation}
The corresponding anomalous Green's function at these three boundaries are wirtten as,
\begin{equation}
F_{(i_y=N_y)}=\frac{\Delta_k t_\perp^2}{(\Delta_k^2+\epsilon_k^2)\omega^2-(-2t_\perp^2+\omega^2)^2}\label{cyyL}.
\end{equation}
\begin{widetext}
\begin{equation}
F_{(i_x=1)}=-\frac{\Delta_k  t_\perp^2 \left(2 k^2+2 k \omega +\omega ^2\right)}{4 k^2 \left(t_\perp^4-\omega ^2 \left(\Delta_k ^2+2 t_\perp^2+\epsilon_k ^2\right)+\omega ^4\right)+4 k \omega  \left(2 t_\perp^4-\omega ^2 \left(\Delta_k ^2+3 t_\perp^2+\epsilon_k ^2\right)+\omega ^4\right)+4 t_\perp^4 \omega ^2-\omega ^4 \left(\Delta_k ^2+4 t_\perp^2+\epsilon_k ^2\right)+\omega ^6}\label{cyxL}
\end{equation}

\begin{equation}
F_{(i_x=N_x)}=-\frac{\Delta_k  t_\perp^2 \left(2 k^2-2 k \omega +\omega ^2\right)}{4 k^2 \left(t_\perp^4-\omega ^2 \left(\Delta_k ^2+2 t_\perp^2+\epsilon_k ^2\right)+\omega ^4\right)+4 k \omega  \left(-2 t_\perp^4+\omega ^2 \left(\Delta_k ^2+3 t_\perp^2+\epsilon_k ^2\right)-\omega ^4\right)+4 t_\perp^4 \omega ^2-\omega ^4 \left(\Delta_k ^2+4 t_\perp^2+\epsilon_k ^2\right)+\omega ^6}\label{cyx1}
\end{equation}
\end{widetext}

At low energies, the momentum $k$ is small. We rewrite the above anomalous Green's functions with the Taylor series Expansions up to the second order approximation, with,

\begin{widetext}
\begin{equation}
\begin{aligned}
F_{(i_y=1)}\approx&-\frac{\Delta_k  t_\perp^2}{-\Delta_k ^2 \omega ^2+4 t_\perp^4-4 t_\perp^2 \omega ^2+\omega ^4-\omega ^2 \epsilon_k ^2}-\frac{4  \left(\Delta_k  t_\perp^2 \left(\Delta_k ^2 \omega +2 t_\perp^2 \omega -\omega ^3+\omega  \epsilon_k ^2\right)\right)}{\left(-\Delta_k ^2 \omega ^2+4 t_\perp^4-4 t_\perp^2 \omega ^2+\omega ^4-\omega ^2 \epsilon_k ^2\right)^2}k\\
&+\frac{4 \Delta_k   t_\perp^2 \left(-4 t_\perp^4 \left(\Delta_k ^2+\epsilon_k ^2\right)+6 \omega ^4 \left(\Delta_k ^2+2 t_\perp^2+\epsilon_k ^2\right)-3 \omega ^2 \left(\Delta_k ^2+2 t_\perp^2+\epsilon_k ^2\right)^2-3 \omega ^6\right)}{\left(4 t_\perp^4-\omega ^2 \left(\Delta_k ^2+4 t_\perp^2+\epsilon_k ^2\right)+\omega ^4\right)^3}k^2,
\end{aligned}
\end{equation}
\begin{equation}
F_{(i_y=N_y)}=-\frac{\Delta_k  t_\perp^2}{-\Delta_k ^2 \omega ^2+4 t_\perp^4-4 t_\perp^2 \omega ^2+\omega ^4-\omega ^2 \epsilon_k ^2},
\end{equation}
\end{widetext}

\begin{widetext}
\begin{equation}
\begin{aligned}
F_{(i_x=1)}\approx&-\frac{\Delta_k  t_\perp^2}{-\Delta_k ^2 \omega ^2+4 t_\perp^4-4 t_\perp^2 \omega ^2+\omega ^4-\omega ^2 \epsilon_k ^2}+\frac{2 \Delta_k   t_\perp^2 \omega  \left(\Delta_k ^2+2 t_\perp^2-\omega ^2+\epsilon_k ^2\right)}{\left(-\Delta_k ^2 \omega ^2+4 t_\perp^4-4 t_\perp^2 \omega ^2+\omega ^4-\omega ^2 \epsilon_k ^2\right)^2}k\\
&+\frac{2 \Delta_k   t_\perp^2 \left(-3 \omega ^4 \left(\Delta_k ^2-\omega ^2+\epsilon_k ^2\right)^2-8 t_\perp^8+24 t_\perp^6 \omega ^2+6 t_\perp^4 \omega ^2 \left(\Delta_k ^2-5 \omega ^2+\epsilon_k ^2\right)-16 t_\perp^2 \omega ^4 \left(\Delta_k ^2-\omega ^2+\epsilon_k ^2\right)\right)}{\omega ^2 \left(4 t_\perp^4-\omega ^2 \left(\Delta_k ^2+4 t_\perp^2+\epsilon_k ^2\right)+\omega ^4\right)^3}k^2,
\end{aligned}
\end{equation}
\begin{equation}
\begin{aligned}
F_{(i_x=N_x)}\approx&-\frac{\Delta_k  t_\perp^2}{-\Delta_k ^2 \omega ^2+4 t_\perp^4-4 t_\perp^2 \omega ^2+\omega ^4-\omega ^2 \epsilon_k ^2}-\frac{2  \left(\Delta_k  t_\perp^2 \omega  \left(\Delta_k ^2+2 t_\perp^2-\omega ^2+\epsilon_k ^2\right)\right)}{\left(-\Delta_k ^2 \omega ^2+4 t_\perp^4-4 t_\perp^2 \omega ^2+\omega ^4-\omega ^2 \epsilon_k ^2\right)^2}k\\
&+\frac{2 \Delta_k   t_\perp^2 \left(-3 \omega ^4 \left(\Delta_k ^2-\omega ^2+\epsilon_k ^2\right)^2-8 t_\perp^8+24 t_\perp^6 \omega ^2+6 t_\perp^4 \omega ^2 \left(\Delta_k ^2-5 \omega ^2+\epsilon_k ^2\right)-16 t_\perp^2 \omega ^4 \left(\Delta_k ^2-\omega ^2+\epsilon_k ^2\right)\right)}{\omega ^2 \left(4 t_\perp^4-\omega ^2 \left(\Delta_k ^2+4 t_\perp^2+\epsilon_k ^2\right)+\omega ^4\right)^3}k^2.
\end{aligned}
\end{equation}

\end{widetext}
The $k^2$ terms are the effective singlet channel pairing terms contributed by the gapless edge modes.
For the $i_x=1$ and $i_x=N_x$ boundaries, the anomalous Green's functions at the singlet channel are exactly the same.
For the $i_y=N_y$ boundary, the contributions from the interorbital hopping and the intraorbital hopping terms cancel out. Thus at this boundary, the effective pairing magnitude is rather small, consistent with our numerical results in the main text.

%
\end{document}